\documentclass[conference]{IEEEtran}

\ifCLASSINFOpdf
 
\else
 
\fi

\usepackage{array}
\usepackage{amsthm,amsmath,amssymb}
\usepackage{mathrsfs}
\usepackage{multirow}
\usepackage{siunitx}

\usepackage{booktabs}
\usepackage{graphicx}
\usepackage{enumerate}
\usepackage{cite}

\begin{document}
\newcommand{\ee}{\end{equation}}
\newcommand{\br}{{\mbox{\boldmath{$r$}}}}
\newcommand{\bp}{{\mbox{\boldmath{$p$}}}}
\newcommand{\bpi}{\mbox{\boldmath{ $\pi $}}}
\newcommand{\bn}{{\mbox{\boldmath{$n$}}}}
\newcommand{\balfa}{{\mbox{\boldmath{$\alpha$}}}}
\newcommand{\ba}{\mbox{\boldmath{$a $}}}
\newcommand{\bta}{\mbox{\boldmath{$\beta $}}}
\newcommand{\bg}{\mbox{\boldmath{$g $}}}
\newcommand{\bPsi}{\mbox{\boldmath{$\Psi $}}}
\newcommand{\bsigma}{\mbox{\boldmath{ $\Sigma $}}}
\newcommand{\bGamma}{{\bf \Gamma }}
\newcommand{\bA}{{\bf A }}
\newcommand{\bP}{{\bf P }}
\newcommand{\bX}{{\bf X }}
\newcommand{\bI}{{\bf I }}
\newcommand{\bR}{{\bf R }}
\newcommand{\bZ}{{\bf Z }}
\newcommand{\bz}{{\bf z }}
\newcommand{\bx}{{\mathbf{x}}}
\newcommand{\bM}{{\bf M}}
\newcommand{\bU}{{\bf U}}
\newcommand{\bD}{{\bf D}}
\newcommand{\bJ}{{\bf J}}
\newcommand{\bH}{{\bf H}}
\newcommand{\bK}{{\bf K}}
\newcommand{\bm}{{\bf m}}
\newcommand{\bN}{{\bf N}}
\newcommand{\bC}{{\bf C}}
\newcommand{\bL}{{\bf L}}
\newcommand{\bF}{{\bf F}}
\newcommand{\bv}{{\bf v}}
\newcommand{\bSigma}{{\bf \Sigma}}
\newcommand{\bS}{{\bf S}}
\newcommand{\bs}{{\bf s}}
\newcommand{\bO}{{\bf O}}
\newcommand{\bQ}{{\bf Q}}
\newcommand{\btr}{{\mbox{\boldmath{$tr$}}}}
\newcommand{\bNSCM}{{\bf NSCM}}
\newcommand{\barg}{{\bf arg}}
\newcommand{\bmax}{{\bf max}}
\newcommand{\test}{\mbox{$
	\begin{array}{c}
		\stackrel{ \stackrel{\textstyle H_1}{\textstyle >} } { \stackrel{\textstyle <}{\textstyle H_0} }
	\end{array}
	$}}
\newcommand{\tabincell}[2]{\begin{tabular}{@{}#1@{}}#2\end{tabular}}
\newtheorem{Def}{Definition}
\newtheorem{Pro}{Proposition}
\newtheorem{Lem}{Lemma}
\newtheorem{Exa}{Example}
\newtheorem{Rem}{Remark}
\newtheorem{Cor}{Corollary}
\renewcommand{\labelitemi}{$\bullet$}

\title{Multi-target Joint Tracking and Classification Using\\ the Trajectory PHD Filter }

\author{Shaoxiu~Wei,
	Boxiang~Zhang,
	and~Wei~Yi \\
		\IEEEauthorblockA{\textit{University of Electronic Science and Technology of China, Chengdu, China} \\
		Email: sxiu\_wei@qq.com, 201811011921@std.uestc.edu.cn, kussoyi@gmail.com}
}

\maketitle

\IEEEpeerreviewmaketitle

\begin{abstract}
	To account for joint tracking and classification
	(JTC) of multiple targets from observation sets in presence of detection
	uncertainty, noise and clutter, this paper develops a new trajectory probability hypothesis density (TPHD) filter, which is referred to as the JTC-TPHD filter. The JTC-TPHD filter classifies different targets based on their motion models and each target is assigned with multiple class hypotheses. By using this strategy, we can not only obtain the category information of the targets, but also a more accurate trajectory estimation than the traditional TPHD filter. The JTC-TPHD filter is derived by finding the best Poisson posterior approximation over trajectories on an augmented state space using the Kullback-Leibler divergence (KLD) minimization. The Gaussian mixture is adopted for the implementation, which is referred to as the GM-JTC-TPHD filter. The $L$-scan approximation is also presented for the GM-JTC-TPHD filter, which possesses lower computational burden. Simulation results show that the GM-JTC-TPHD filter can classify targets correctly and obtain accurate trajectory estimation.
\end{abstract}

\IEEEpeerreviewmaketitle

\section{Introduction}
\IEEEPARstart
{T}{he} probability hypothesis density (PHD) filter~\cite{Mahler2007RFSbook,Mahler2003RFS,Vo2006PHD} is a widely used approach in the multi-target tracking, which aims to model the appearance and disappearance of targets, false detections and misdetections of measurements based on the random finite set (RFS)~\cite{Mahler2007RFSbook}. The PHD filter is known for its low computational burden, which considers a Poisson multi-target filtering density. If the prior or posterior density is not
Poisson, the PHD filter finds the best Poisson approximation to enjoy a conjugate closure by minimizing the Kullback-Leibler divergence (KLD) \cite{Angel2015KLD}. There are several common implementations for the PHD filter, such as sequential Monte Carlo (SMC) \cite{Vo2005smc} or Gaussian mixture (GM) \cite{Vo2006PHD}. Following the same routine, the cardinalized PHD
(CPHD) filter \cite{Vo2007CPHD}, the multi-Bernoulli (MB) filter \cite {Vo2009MB}, the Poisson multi-Bernoulli mixture (PMBM) filter \cite{Angel2018PMBM}, the generalized labeled multi-Bernoulli (GLMB) \cite{Vo2013GLMB, Vo2014GLMB}
filter, and the labeled multi-Bernoulli (LMB) filter \cite{ Reuter2014LMB} are also devised.

\par In the multi-target tracking, we also need to obtain more accurate trajectory estimation and eliminate the
trajectory fragmentation \cite{Vo2019MsGLMB}. Recently, using a set of trajectories as the posterior density ~\cite{Angel2019TPHD,Granstrom2018TPMBM,Angel2020TMB,Angel2020TMTT}
or a (labeled) multi-target state sequence posterior \cite{Vo2019MsGLMB} provides
an efficient approach to the requirements above. Among these approaches, the trajectory PHD (TPHD) filter \cite{Angel2019TPHD} establishes trajectories from first principles using trajectory
RFSs. The TPHD filter propagates the best Poisson multi-trajectory approximation using the KLD minimization \cite{Angel2015KLD}. The Gaussian mixture is proposed to obtain a closed-form solution of the TPHD filter, which is given as the GM-TPHD filter. Meanwhile, the $L$-scan approximation strategy is suggested to
achieve the fast implementation of the TPHD filter by only updating the multi-trajectory density of the last $L$ and keeps the rest unchanged.

\par The joint tracking and classification (JTC) \cite{Yw2012JTC, Magnant2016MSPHD, Lsun2014JTC, Challa2001JTC-ESM, JTC-GLMB1, JTC-GLMB2,JTC1} of targets is also a critical
problem in radar detection fields. For example, in the battlefield surveillance, we need to identify incoming aircrafts and missiles instead of only considering tracking them. Generally, the motion models \cite{Yw2012JTC} or other characteristics of targets like the extended target model \cite{Lsun2014JTC} are related to their categories. In principle, correctly classifying targets and assigning a class-matched filter can also improve the accuracy of both detection and track estimation. In \cite{Yw2012JTC}, the JTC model is extended to the PHD filter with the particle implementation, which considers each target class is assigned with a class-matched PHD-like filter. This filter propagates particles based on 
their class-dependent motion models in the prediction step and exchanges
the mutual information between these PHD-like filters
by updating the particle weights. Later, in \cite{Magnant2016MSPHD}, the Gaussian implementation of the jump Markov system \cite{Pasha2009JMS, MM1, MM2, MM3} PHD (JMS-PHD) filter combined with the JTC model is discussed. Both above JTC methods are based on the estimation of the mixed density probability, and the classification is significantly dependent on the estimation. It is worth noting that, in \cite{JTC-GLMB2}, the GLMB filter is combined with a novel joint decision and estimation algorithm \cite{CJDE1,CJDE2}
based on the generalized Bayes risk to solve the JTC problem. Such risk defined in the GLMB filter involves both the estimation 
costs of cardinality and states, and the classification cost. 

\par In this paper, the concept of JTC is developed into the trajectory RFS.
Combined with the JTC model, a new TPHD filter is proposed to obtain the trajectory information and classify different kinds of targets, which is referred to as the JTC-TPHD filter. In this filter, we assign possible category hypotheses for each trajectory and the corresponding multiple motion models for each category. After the prediction and update step, the category hypothesis with the maximum posterior probability is extracted to serve as the classification result of the target at this moment, and then the trajectory estimation of the target is extracted from this category hypothesis. Besides, the Gaussian mixture is adopted to obtain an efficient implementation of the JTC-TPHD filter, named as the GM-JTC-TPHD filter. The $L$-scan approximation of the GM-JTC-TPHD filter is
adopted to deal with the huge computational burden caused
by the increasing length of trajectory. Simulation results demonstrate the GM-JTC-TPHD filter can achieve excellent performance in tracking and classification.

\section{Background}
This section provides a brief review of the TPHD filter. The notation instruction is given in section II-A. The Bayesian filtering recursion for trajectories is elaborated in section II-B. Finally, the recursion of the TPHD filter is given section II-C. Further details can be found in \cite{Angel2019TPHD}.
\subsection{Sets of Trajectories}
The trajectory state $X=(\beta,x^{1:\zeta})$ consists of a finite sequence
of target states $x^{1:\zeta}=(x^1,...,x^\zeta)$ that is born at time step $\beta$ with length $\zeta$ \cite{Angel2019TPHD}. For a trajectory $(\beta,x^{1:\zeta})$ that exists from time $\beta$ to $\beta+\zeta-1$, the variable $(\beta, \zeta)$ belongs to the set $\mathbb{U}_k=\{(\beta, \zeta):1\leq \beta\leq k~\text{and}~1\leq \zeta\leq k-\beta+1\}$. Therefore, a single trajectory up to time step $k$
belongs to the space $\mathbb{T}_k=\uplus_{(\beta, \zeta)\in \mathbb{U}_k}\{\beta\}\times\mathbb{R}^{\zeta\times n_x}$, where $\uplus$ denotes the disjoint union, $\times$ denotes a Cartesian product, and $n_x$ represents the dimension of target state. Supposing there are $N^k$ trajectories at time $k$, the set of trajectories is denoted as 
\begin{align}
\mathbf{X}_k=\{X_{1},...,X_{N^k}\}\subset \mathcal{F}(\mathbb T_k),
\end{align} 
where $\mathcal{F}(\mathbb{T}_k)$ is the respective collections of all finite subsets of $\mathbb T_k$. The inner product between two real valued functions $a$ and $b$ is $\langle {a,b}\rangle$, which equals to $\int {a( x )b(x)} dx$. The generalized Kronecker delta function is expressed as,
\begin{align}
\delta_{A}(B) \triangleq \left\{
\begin{array}{lr}
	1, ~~\text{if}~B=A& \\
	0, ~~\text{otherwise},&
\end{array}
\right.
\end{align}
and the inclusion function is given as
\begin{align}
	1_{Y}(X) \triangleq \left\{
	\begin{array}{lr}
		1, ~~\text{if}~X\subseteq Y& \\
		0, ~~\text{otherwise}.&
	\end{array}
	\right.
\end{align}
\subsection{Bayesian Filtering Recursion}
Given the posterior multi-trajectory density $\pi_{k-1}(\cdot)$ at time $k-1$ and the set of measurements $\mathbf{z}_k$ at time $k$, the posterior density $\pi_{k}(\cdot)$ can be obtained by using the Bayes recursion
\begin{align}
	\pi_{k|k-1}\left(\mathbf{X}_k \right) =& \int {f\left( {\mathbf{X}_k |\mathbf{X}_{k-1} } \right)} {\pi_{k - 1}}\left( \mathbf{X}_{k-1} \right)\delta \mathbf{X}_{k-1} , \\
	\pi_k\left( \mathbf{X}_k \right) =& \frac{{{\ell_k}\left( {\mathbf{z}_k|\mathbf{X}_k} \right){\pi_{k|k - 1}}\left( \mathbf{X}_k \right)}}{	\int {{\ell_k}\left( {\mathbf{z}_k| \mathbf{X}_k} \right)} {\pi_{k|k - 1}}\left( \mathbf{X}_k \right)\delta \mathbf{X}_k},
\end{align}
where $f( { \cdot | \cdot })$ denotes the transition density, $ \pi_{k|k - 1}(\cdot)$ denotes the predicted density, ${\ell_k}\left(\mathbf{z}_k|\mathbf{X}\right)$ denotes the density of measurements of trajectories. As the measurements come from the target states of a single frame time, ${\ell_k}\left( \mathbf{z}_k|\mathbf{X}_k \right)$ can be also written as
\begin{equation}
	{\ell_k}\left( {\mathbf{z}_k|\mathbf{X}_k} \right) = {\ell_k}\left( {\mathbf{z}_k|{\tau _k}\left( \mathbf{X}_k \right)} \right),
\end{equation}
where $\tau_k(\mathbf X)$ denotes the corresponding multi-target state at the time $k$. Similarly, the measurement likelihood $l_{k}(z|X)$ of the single trajectory $X=(\beta,x^{1:\zeta})$ equals to $l_{k}(z|x^\zeta)$. 
\subsection{The TPHD Filter}
\par The TPHD filter considers a Poisson multi-trajectory density $v_k(\cdot)$ at time $k$, which is expressed as
\begin{equation}
	p_k(\{X_1,...,X_{N^k}\})=e^{-\lambda_k}\lambda_k^n\prod_{j=1}^{N^k}\bar{p}_k(X_j),
\end{equation}
where $\bar{p}_k(\cdot)$ denotes the single trajectory density and $\lambda_k\ge0$. A Poisson PDF is characterized by its PHD \cite{Angel2015KLD}
\begin{align}\label{PHD-trajectory}
	D_k(X) = \lambda_k\bar{p}_k(X).
\end{align}
The clutter RFS is also Poisson with intensity $\kappa(z)$. The TPHD filter follows the assumptions \cite{Angel2019TPHD}:
\begin{itemize}
	\item [$\bullet$]
	 The trajectories st time $k$ consist of surviving trajectories at time $k-1$ with surviving probability ${p_{S,k}}\left(  \cdot  \right)$, and the trajectories born at time $k$ with the PHD ${{\gamma_k }}( \cdot)$ of a Poisson density. The birth and the surviving RFSs are independent of each other.
	\item [$\bullet$]
	The trajectory RFS at time $k-1$ is Poisson and the clutter RFS is also Poisson and independent of measurement RFS.
\end{itemize}  
\par Given the posterior PHD $D_{k-1}(\beta,x^{1:\zeta-1})$ at time $k-1$ and the transition density $f( {x^\zeta|x^{\zeta-1}})$, the prediction step of the TPHD filter is obtained by
\begin{equation}
	{D_{k|k - 1}}\left( X \right) = {\gamma _k}\left( X \right) + D_k^S \left( X \right),
\end{equation}
where
\begin{align}
	{\gamma _k}\left( X \right) &= \gamma_k \left( {k,{x^1}} \right),\\
	D_k^S \left( X \right) &= {p_{S,k}}\left( x^\zeta \right)f\left( {x^\zeta|x^{\zeta-1}} \right){D_{k - 1}}\left( \beta,x^{1:\zeta-1} \right).\label{pr_TPHD}
\end{align}
\par As only alive trajectories are considered in \cite{Angel2019TPHD}, if $\zeta\neq k-\beta+1$, $D^S_k(X)$ equals to zero. Besides, in (\ref{pr_TPHD}), it is required $\beta\in\{1,2,...,k-1\}$ to represents the trajectories born before time $k$. The update step of the TPHD filter is obtained by
\begin{align}
	{D_k}\left( X \right) =& {D_{k|k - 1}}\left( X \right)({1-p_{D,k}\left( x^\zeta \right)})\\
	&+{D_{k|k - 1}}\left( X \right)p_{D,k}(x^\zeta)\notag\\
	&\times\sum\limits_{z \in {Z_k}} {\frac{{{l_k}(z|x^\zeta)}}{{ \kappa(z) + \left\langle {p_{D,k}\cdot{l_k}\left( {z| \cdot } \right),D_{k|k-1}^\tau } \right\rangle }}}\notag,
\end{align}
where
\begin{equation}
	D_{k|k-1}^\tau \left( {x^\zeta}\right)=\sum\limits_{\beta = 1}^k {\int {{D_{k|k - 1}}(t,{x^{1:k - \beta +1}})d{x^{1:k - \beta }}} },
\end{equation}
with $\zeta=k-\beta+1$,
which denotes the PHD of the prior target density at time $k$. 
\section{The JTC-TPHD Filter}
In this section, the recursion of the JTC-TPHD filter is elaborated. The JTC-TPHD filter can classify the target when tracking and each target is assigned with multiple class hypotheses. Using this strategy, we can obtain both the category information of the targets and a more accurate trajectory estimation than the traditional TPHD filter. In Section III-A, the JTC model is presented. In Section III-B, the JTC-TPHD filter is derived in detail. 
\subsection{The Joint Tracking and Classification Model}
Assume that a considered target possesses the class hypothesis $c \in \mathbb{C}_k$ at time $k$, where $\mathbb{C}_k$ represents the state space of classes. Each target class is then characterized
by a finite number of possible motion models denoted by $r \in \mathbb{M}(c)$, where $r$ denotes the possible motion model and $\mathbb{M}(c)$ denotes the corresponding space of motion models with class $c$. 
The class state is augmented to the trajectory state, and the state space of the augmented trajectories is defined as
\begin{align}
\mathbb{Y}=\uplus_{(\beta,\zeta)\in \mathbb{U}_k}\{\beta\}\times\mathbb{R}^{\zeta\times n_x}  \times \mathbb C,
 \end{align}
where $\mathbb Y$ represents the space of the augmented trajectories ${X_c}$, which consists of trajectories ${X}$ and the corresponding classes ${c}$, i.e. ${X_c}=(X,c)$. For a single trajectory $X$ at time $k$, the classification result is determined by the category hypothesis with the maximum posterior probability density at this moment. Given the sets of measurements $\mathbf{z}_{1:k}$
from time $1$ to $k$, the prior density $Pr(c|\mathbf{z}_{1:k-1})$ and the likelihood function $p(\mathbf{z}_k|\cdot)$ of of class $c$, the posterior probability density is computed by the Bayes rule
\begin{align}
	Pr\left( c|\mathbf{z}_{1:k} \right) =& \frac{p(\mathbf{z}_k|c,\mathbf{z}_{1:k-1})Pr(c|\mathbf{z}_{1:k-1})}{\sum_{c' \in \mathbb{C}_k}p(\mathbf{z_k}|c',\mathbf{z}_{1:k-1})Pr(c'|\mathbf{z}_{1:k-1})}.
\end{align} Besides, we define the density of augmented trajectory sets as the augmented multi-trajectory density. The augmented trajectory RFS is denoted as
\begin{align}
	\mathbf{X_c}=\{X_c=(\beta,x^{1:\zeta},c^\zeta)\in \mathbb {Y}\}.
\end{align}
More specifically, considering an augmented trajectory possesses a motion model $r^\zeta$, then the state is expressed as
\begin{align}
	X_{c,r}=(\beta,x^{1:\zeta},c^\zeta,r^\zeta)\in \mathbb Y \times \mathbb{M}(c^\zeta),
\end{align}
For an augmented trajectory $X_{c,r}=(\beta,x^{1:\zeta},c^\zeta,r^\zeta)$ with a motion model at time $k$, its transition density is given as ${\hat f} (X_{c,r}|\tilde X_{c,r})$, where $\tilde X_{c,r}=(\underline\beta,\underline{x}^{1:\underline{\zeta}},c^{\underline{\zeta}},r^{\underline{\zeta}})$ represents the state at time $k-1$.
In some applications, the switch of motion models are independent of the trajectory states, which is given as $f_r(r^\zeta|r^{\zeta-1})$. However, for a certain target, its class does not change with time. In other words, there is no transition between different target classes without considering spawning targets. Besides, the transition ${\hat f} (X_{c,r}|\tilde X_{c,r})$ is considered as the first-order Markov process, thus the following equation is established
\begin{align}
	{\hat f}(X_{c,r}|\tilde X_{c,r})=&f\left( x^\zeta|x^{\zeta-1},r^{\zeta-1}\right)f_r\left( r^\zeta|r^{\zeta-1} \right)\delta_{c^{\underline\zeta}}[c^{\zeta}]\notag\\
	&\times\delta_{\underline{x}^{1:\underline{\zeta}}}({x}^{1:\zeta})
	\delta_{\underline{\beta}}[{\beta}]\delta_{\underline{\zeta}+1}[{\zeta}],
\end{align}
In this paper, the measurements are only concerned about
target states of a single frame time, thus the measurement likelihood $l_k(z|X_{c,r})$
can be simplified to
\begin{align}
	l_k(z|X_{c,r})=l_k(z|x^\zeta).
\end{align}
\subsection{The JTC-TPHD Filter}
\par Following the work of the JTC-PHD fillter \cite{Yw2012JTC}, the recursion of the JTC-TPHD filter not only extends the JTC model into trajectory RFS, but also presents the prediction and update process of the categories and multiple motion models of targets in detail. The JTC-TPHD filter follows the same assumptions as the TPHD filter \cite{Angel2019TPHD}, which is not elaborated here. Combined with JTC model, the PHD of the multi-trajectory density \eqref{PHD-trajectory} at time $k$ is expressed as,
\begin{align}
	D_k(X)=\sum\limits_{c^\zeta\in\mathbb C}D_{c,k}(X_c)=\sum\limits_{c^\zeta\in\mathbb C}\sum\limits_{r^\zeta\in\mathbb{M}(c^\zeta)}D_{r,k}(X_{c,r}),
\end{align}
where $D_{c,k}(X_c)$ and $D_{r,k}(X_{c,r})$ represent the PHD of the augmented multi-trajectory density considering class and the PHD of the augmented multi-trajectory density with motion model $r$, respectively. By propagating the PHD $D_c(X_c)$ of the augmented trajectory $X_c=(\beta,x^{1:\zeta},c^{\zeta})$, the recursion of the JTC-TPHD filter is given as follows,

\begin{Pro}\label{JTCTPHD_Pr}
	Given the posterior PHD $D_{c,k-1}(X_c)$ at time $k-1$, which equals to
	\begin{align}
	{D_{c,k - 1}}&\left(\beta,x^{1:\zeta-1},c^{\zeta-1}\right)\\
	&=\sum\limits_{r^{\zeta-1}\! \in \mathbb{ M}(c^{\zeta-1})}\!\!\!{D_{r,k - 1}}\left(\beta,x^{1:\zeta-1},c^{\zeta-1},r^{\zeta-1}\right),\notag 
	\end{align}
	 the predicted PHD $D_{c,k|k-1}(X_c)$ of the augmented multi-trajectory density is obtained by
\end{Pro}
\begin{align}
{D_{c,k|k - 1}}\left(X_c \right)&= {\gamma _{c,k}}\left( X_c \right) + D_{c,k}^S \left( X_c \right),
\end{align}
where
\begin{align}
	{\gamma _{c,k}}\left( X_c \right) =&  \gamma _{c,k} \left( {k,{x^1},c^1} \right),\\
	D_{c,k}^\zeta \left( X_c \right) =&  \delta_{c^{\zeta-1}}(c^{\zeta})\sum_{r^\zeta\in\mathbb{M}(c^\zeta)}\sum_{r^{\zeta-1}\in\mathbb{M}(c^{\zeta-1})} f_r\left( {r^\zeta|r^{\zeta-1}}\right)\notag\\
	&\times p_{S,k}\left( x^{\zeta-1},c^{\zeta-1}\right)f\left( {x^\zeta|x^{\zeta-1},r^{\zeta-1}} \right) \label{pr_TPHD_s}\\
	&\times{D_{r,k - 1}}\left( \beta,{x^{1:\zeta - 1}},c^{\zeta - 1},r^{\zeta - 1} \right),\notag	
\end{align}
\par As indicated by (\ref{pr_TPHD_s}), for surviving trajectories, the prediction step includes the switches of different target motion models of the same class and the prediction of trajectory states. Note that, different from \cite{Yw2012JTC}, the previous states of trajectories are retained in (\ref{pr_TPHD_s}).
\begin{Pro}\label{JTCTPHD_Up}
	Given the predicted PHD $D_{c,k|k-1}(X_c)$ at time $k$, the posterior PHD $D_{c,k}(X_c)$ of the multi-trajectory density is obtained by
\end{Pro}
\begin{align}
	{D_{c,k}}\left( \beta,x^{1:\zeta},c^\zeta \right) =& D_{c,k|k - 1}\left( \beta,x^{1:\zeta},c^\zeta \right)(1-p_D(x^\zeta,c^\zeta))\notag\\
	& + D_{c,k|k - 1}\left(\beta,x^{1:\zeta},c^\zeta \right)p_D(x^\zeta,c^\zeta)\notag\\
	&\times\sum\limits_{z \in {Z_k}} {\frac{l_k(z|x^\zeta)}{\kappa(z)+\Theta_k\left[z,X_c\right]}},
\end{align}
where
\begin{align}
	\Theta_k\left[z,X_c\right]=&\sum_{c^\zeta\in\mathbb{C}}\int p_D(x^\zeta,c^\zeta)\cdot l_k(z|x^\zeta)\\
	&\times D^\tau_{c,k|k-1}(x^\zeta,c^\zeta)dx^\zeta,\notag\\
		D_{c,k|k-1}^\tau(x^\zeta,c^\zeta)=&\sum\limits_{\beta = 1}^k\int {D_{c,k|k - 1}(\beta,x^{1:k-\beta+1},c^{k-\beta+1})}\notag\\
		&{\times dx^{1:k-\beta}}.
	\end{align}
\par In Proposition \ref{JTCTPHD_Up}, $\zeta=k-\beta+1$, and $D_{c,k|k-1}^\tau (x^\zeta,c^\zeta)$ denotes the PHD of the prior density of augmented targets at time $k$. The updated PHD contains information about the trajectories, corresponding classes and motion models.
\section{The Gaussian Mixture Implementation}
In this section, the Gaussian mixture is presented to obtain a  closed-form implementation for the JTC-TPHD filter, which is referred to as the GM-JTC-TPHD filter. There are some notations given as follows. At time $k$, the Gaussian density of a trajectory which is born at time $t$ of length $i$ is given as
\begin{align}
	{\cal N}( {X_{c,r};\beta,\widehat{m}^k(r),\widehat{P}^k(r)}) &= {\cal N}( {{x^{1:\zeta}};\widehat{m}^k(r),\widehat{P}^k(r)})\\
	&\times1_{\mathbb{M}(c)}(r)1_{\mathbb{U}_k}[\beta,\zeta]1_{\mathbb{C}_k}(c),\notag
\end{align}
with the mean $\widehat{m}^k \in \mathbb{R}^{\zeta\times{n_x}}$ and the covariance $\widehat{P}^k \in \mathbb{R}^{\zeta{n_x}\times \zeta{n_x}}$. For a matrix $V$, the notation ${V_{\left[ {n:m,s:t} \right]}}$ represents the submatrix of $V$ for rows from time steps $n$ to $m$ and columns from time steps $s$ to $t$. The notation $V_{[n:m]}$ is used to present the submatrix of $V$ for rows from time steps $n$ to $m$. Besides, the notations $V_{[n,s:t]}$ and $V_{[n]}$ represent $V_{[n:n,s:t]}$ and $V_{[n:n]}$, respectively. The recursion of the GM-JTC-TPHD filter follows the assumptions of
\par\emph{Assumption 1:} The transition of target state and observation function are the linear Gaussian model
\begin{align}
	f\left( x^\zeta|x^{\zeta - 1},r^{\zeta - 1}\right)=&{\cal N}\left( {x^\zeta};F(r^{\zeta}){x^{\zeta - 1}},Q(r^{\zeta})\right),\\
	l\left( {z|x^\zeta} \right)=&{\cal N}\left( {z;Hx^\zeta,R} \right), \label{Gaussian}
\end{align}
where $F \in \mathbb{R}^{n_x \times n_x}$ denotes the state transition matrix and $Q\in \mathbb{R}^{n_x \times n_x}$ is the process noise covariance.
$H \in \mathbb{R}^{n_z \times n_x}$ is the observation matrix and $R \in \mathbb{R}^{n_z \times n_z}$
denotes the observation noise covariance.
\par\emph{Assumption 2:} The values of surviving probability and detection probability are considered as class dependent in the implementation, which are expressed as, $p_S (c)$ and $p_D (c)$, respectively.
\par \emph{Assumption 3:}  The PHD of the birth density $\gamma _{c,k}$ at time $k$ is a Gaussian mixture
\begin{align}
	{\gamma _{c,k}}\left( X_c \right) =& \sum\limits_{j = 1}^{J_\gamma ^k(c^k)} \omega _{\gamma_c ,j}^k(c^k)\sum_{r^k\in\mathbb{M}(c^k)}\omega _{\gamma_r ,j}^k(r^k)\\
	&\times{\cal N}\left( {X_{c,r};k,\widehat m_{\gamma ,j}^k(r^k),\widehat P_{\gamma ,j}^k(r^k)} \right),\notag
\end{align}
where $J^k_\gamma \in \mathbb{N}$ represents the number of Gaussian components. For the $j$-{th} birth component at time $k$, $\omega_{\gamma_c,j}^k$ and $\omega_{\gamma_r,j}^k$ represents the weight of components and motion models, respectively. The mean and covariance of the Gaussian density are expressed as ${\widehat m}_{\gamma,j}^k \in \mathbb{R}^{n_x}$ and ${\widehat P}_{\gamma,j}^k\in \mathbb{R}^{n_x \times n_x}$, respectively. 
\begin{Pro}\label{UBGTPHD_Pr}
	The posterior PHD ${D_{c,k - 1}}( {X_c})$ at time $k-1$ can be given as the Gaussian mixture as follows
	\begin{align}
		{D_{c,k - 1}}&\left( X_c \right) = \sum\limits_{j = 1}^{J^{k - 1}(c^{k-1})} \omega _{c,j}^{k - 1}(c^{k-1})\!\!\!\sum\limits_{r^{k-1}\in\mathbb{M}(c^{k-1})}\!\omega _{r,j}^{k - 1}(r^{k-1})\notag\\
		&\times{\cal N}\left(X_{c,r};\beta_j,\widehat{m}_j^{k-1}(r^{k-1}),\widehat{P}_j^{k-1}(r^{k-1}) \right),
	\end{align}
	where, at time $k-1$, the length $\zeta^{k-1}_j$ of the $j$-{th} augmented trajectory $X_{c,r}$ is given as $k-\beta_j$. The mean and covariance of the Gaussian density are given as $\widehat m_j^{k-1}(r^{k-1})=\in \mathbb{R}^{\zeta^{k-1}_j\times n_x }$ and $\widehat P_j^{k-1}(r^{k-1})\in \mathbb{R}^{\zeta^{k-1}_jn_x \times \zeta^{k-1}_jn_x }$, respectively. 
	Then the prior PHD $D_{c,k|k-1}(X_c)$ is given as
\end{Pro}
\begin{align}
	D_{c,k|k-1}(X_c)=&\gamma _{c,k}(X_c)+D^S_{c,k}(X_c),\\
	{D^S_{c,k}}\left( X_c\right)=&\sum_{r^k\in\mathbb{M}(c^k)}\!\!\!\!\!\!\!\!\sum\limits_{j = 1}^{{J^{k - 1}(c^{k-1})}}\!\!\!\!\!p_S(c^{k-1})\delta_{c^{k-1}}(c^{k})\omega _{c,j}^{k - 1}(c^{k-1})\notag\\ &\times\sum_{r^{k-1}\in\mathbb{M}(c^{k-1})}w_{r,j}^{k-1}(r^{k-1})f_r(r^k|r^{k-1})\notag\\
	&\times{\cal N}\left( {X_{c,r};\beta_j,\widehat{m} _{s,j}^{k|k-1}(r^k),\widehat{P} _{s,j}^{k|k - 1}}(r^k) \right),
\end{align}
where
\begin{align}	
	\widehat m_{s,j}^{k|k - 1}(r^k) =& \left[ {\begin{array}{*{20}{c}}
			{\widehat m_j^{k - 1}(r^{k-1})}\\
			{F(r^{k}) \widehat m_{j,[k-1]}^{k - 1}(r^{k-1})}
	\end{array}} \right],\\
	\widehat P_{s,j}^{k|k - 1}(r^k) =& \left[ {\begin{array}{*{20}{c}}
			{\widehat P_j^{k - 1}(r^{k-1})}&{P_1}\\
			{P_1^\top}&{P_2}
	\end{array}} \right],\\
	P_1=&\widehat P_{j,\left[ {\beta_j:k - 1,k - 1} \right]}^{k - 1}(r^{k-1}){F(r^k)^\top},\\
	P_2=&F(r^k)\widehat P_{j,[k-1,k-1]}^{k - 1}(r^{k-1})F(r^k)^\top + Q(r^k).
\end{align}
\par Propositions 3 is the consequence of Proposition
1. The prediction step of trajectory in the GM-JTC-TPHD filter roots in the change of the target state and the prediction of motion models depends on the transition density function $f_r(\cdot|\cdot)$.
\begin{Pro}\label{GMJTPHD_Up}
	If at time $k$, the prior PHD $D_{c,k|k-1}(X_c)$ 
	is given as the Gaussian mixture of the form
	\begin{align*}
		{D_{c,k|k - 1}}\left( X_c \right) &= \sum\limits_{j = 1}^{J^{k|k - 1}(c^{k})} \omega _{c,j}^{k|k - 1}(c^{k})\!\!\!\sum\limits_{r^{k}\in\mathbb{M}(c^{k})}\!\omega _{r,j}^{k|k - 1}(r^{k})\\
		&\times{\cal N}\left(X_{c,r};\beta_j,\widehat{m}_j^{k|k-1}(r^k),\widehat{P}_j^{k|k-1}(r^k) \right),
	\end{align*}
	then, given a measurement set $Z_k$, the posterior PHD $D_{c,k}(X_c)$ is given as
\end{Pro}
\begin{align}
	D_{c,k}\left( X_c \right)=& (1-p_D(c^k))D_{c,k|k-1}(X_c)\\
	&+\sum\limits_{z \in {Z_k}} \sum\limits_{j = 1}^{J^{k|k-1}(c^k)} {\omega _{c,j}^k(c^k,z)}\sum\limits_{r^k\in\mathbb{M}(c^k)}\!\!\!\omega _{r,j}^k(r^k,z)\notag\\
	&\times{\cal N}\left( {X_{c,r};\beta_j,\widehat m_j^k\left( r^k,z \right),\widehat P_j^k(r^k)} \right), \notag
\end{align}
where
\begin{align}
	\omega _{r,j}^k(r^k,z) =& \frac{\omega _{r,j}^{k|k-1}(r^k,z)q_j(r^k,z)}{\phi_j(c^k,z)},\\
	\omega _{c,j}^{k}(c^k,z)=&\frac{p_D(c^k)\omega _{c,j}^{k|k - 1}(c^k)\phi_j(c^k,z)}{\kappa(z) + \sum\limits_{l = 1}^{J^{k|k - 1}}\sum\limits_{c^k \in\mathbb{C}} {p_D(c^k)\omega _{c,l}^{k|k - 1}\phi_l(c^k,z)} },\\
	\phi_j(c^k,z)=&\sum_{r^k\in\mathbb{M}(c^k)}\omega _{r,j}^{k|k-1}(r^k,z)q_j(r^k,z),\\
	{\bar z _j(r^k)} =& {H}\widehat m_{j,[k]}^{k|k - 1}(r^{k}),\\
	{S_j}(r^k) =& {H}\widehat P_{j,[k,k]}^{k|k - 1}(r^k)H^\top + R,\\
	q_j(r^k,z)=&{\cal N}(z;\bar z_j(r^k),S_j(r^k)),\\
	\widehat{P} _j^k(r^k) =& \widehat{P}_j^{k|k - 1}(r^k) - {K_j}{H}\widehat{P}_{j,\left[ {k,\beta_j:k} \right]}^{k|k - 1}(r^k),\\
	{K_j} =& \widehat{P} _{j,\left[ {\beta_j:k,k} \right]}^{k|k - 1}(r^k)H^\top (S_j(r^k))^{ - 1}.
\end{align}
\par Different from the JTC-PHD filter \cite{Yw2012JTC}, the GM-JTC-TPHD filter aims at the whole trajectory in the update step. It not only updates the estimation of the target state at current time, but also smooths the estimation of previous states. When there is only one case of the target class $c$, the GM-JTC-TPHD filter degrades into the GM-JMS-TPHD filter \cite{JMS-TPHD}, which only considers tracking maneuvering targets without classification. 
\par Similar to the traditional GM-TPHD \cite{Angel2019TPHD} filter, the number of Gaussian
components for the GM-JTC-TPHD filter increases as time
progresses. Hence, to limit unbounded Gaussian
components, the pruning and absorption techniques are needed, which can be found in \cite{Angel2019TPHD}.
In addition, the $L$-scan approximation \cite{Angel2019TPHD} is also needed to deal with the increasing length of trajectories, which only updates the multi-trajectory density of the last $L$ time and leaves the rest unaltered. At the end of each recursion, the classification result of the $j$-th target is determined by
\begin{align}
	\mathop{max}\limits_{c\in \mathbb{C}_k}~\omega^k_{c,j}(c^k),
\end{align}
then the estimation of the number of alive trajectories of all possible classes at time $k$ is given as
\begin{align}
	{N^k} =  \text{round}( {\sum_{j = 1}^{J^k}\sum_{c^k\in\mathbb{C}_k} {\omega _{c,j}^k}(c^k) }).
\end{align}
The state estimations of $N^k$ trajectories are given as $
\left\{ {\left( {{\beta_1},\zeta_1^k,\widehat m_1^k} \right),...,\left( {{\beta_{{N^k}}},\zeta_{{N^k}}^k,\widehat m_{{N^k}}^k} \right)} \right\}.
$
\section{Simulations}
This section presents numerical studies for the GM-JTC-TPHD filter. A six targets simulation is set inside of a three-dimensional space with the size of $\left[ \rm{-8000},\rm{12000} \right]m\times\left[ {{\rm{-12000}},{\rm{6000}}} \right]m\times\left[ {{\rm{ 0}},{\rm{10000}}} \right]m$  for 100 seconds. The target state matrix is given as $x_k = [p_x,{{\dot p}_x},{p_y},{{\dot p}_y},p_z,{{\dot p}_z}]$, where $p_x,p_y,p_z$ denote the position information and $\dot p_x,\dot p_y,\dot p_z$ represent the velocity information. Consider an nonlinear observation process 
\begin{align}	
	z_k=h_k(x_k,v_k)=&\left[ {\begin{array}{*{20}{c}}
			{\text{tan}^{-1}\left(\frac{p_{x,k}}{p_{y,k}}\right)}\\
			{\text{tan}^{-1}\left(\frac{\sqrt{r_1}}{p_{z,k}}\right)}\\
			{\sqrt{r_2}}
	\end{array}} \right]+v_k, 
\end{align}
where $v_k\sim {\cal N}(\cdot;0,R_k)$ and $R_k=\text{diag}([(\pi/180)^2,(\pi/180)^2,10]^\top)^2$. By using the extended Kalman (EK) \cite{KF-1970} filter to local linearizations of the (nonlinear) mapping $h_k$, we can obtain an approximate linear observation matrix \eqref{Gaussian}
\begin{align}
	H=&\frac{\partial h_k(x_k,0)}{\partial x_k}\\\notag=&\left[ {\begin{array}{*{20}{c}}
			\frac{p_{y,k}}{r_1}&0&\frac{-p_{x,k}}{r_1}&0&0&0\\
			\frac{p_{x,k}p_{z,k}}{r_2\sqrt{r_1}}&0&	\frac{p_{y,k}p_{z,k}}{r_2\sqrt{r_1}}&0&\frac{-\sqrt{r_1}}{r_2}&0\\
			\frac{p_{x,k}}{r_2}&0&\frac{p_{y,k}}{r_2}&0&\frac{p_{z,k}}{r_2}&0
	\end{array}} \right],
\end{align}
where $r_1=p^2_{x,k}+p^2_{y,k}$ and $r_2=p^2_{x,k}+p^2_{y,k}+p^2_{z,k}$.
The targets are divided into two classes: the plane ($c_p$) and unmanned aerial vehicle (UAV) ($c_u$). In other words, the space of class is given as $\mathbb{C}=\{c_p,c_u\}=2$. In this scenario, the plane goes into the nosedive from the high, belonging to the weak maneuvering target and the UAV flies at a low altitude, belonging to the high maneuvering target. The maneuvering motion model consists of the CT and CV modes. The linear state transition matrices for the CV and CT models are given as follows
\begin{align}
\label{equ_cv}	F_{CV} =& I_3\otimes \left[ {\begin{array}{*{20}{c}}
			{1}&{\delta t}\\
			{0}&{1}
	\end{array}} \right],\\
\label{equ_ct}F_{CT} =& \left[ {\begin{array}{*{20}{c}}
		1&\delta t&0&0&0&0\\
		0&1&0&0&0&0\\
		0&0&{1}&{\frac{sin\zeta\delta t}{\zeta}}&{ 0}&{-\frac{1-cos\zeta\delta t}{\zeta}}\\
		0&0&{0}&{cos\zeta\delta t}&{0}&{-sin\zeta\delta t}\\
		0&0&{0}&{\frac{1-cos\zeta\delta t}{\zeta}}&{1}&{\frac{sin\zeta\delta t}{\zeta}}\\
		0&0&{0}&{sin\zeta\delta t}&{0}&{cos\zeta\delta t}
\end{array}} \right],\\
Q_{CV}=&Q_{CT} = \sigma _v^2I_3\otimes\left[ {\begin{array}{*{20}{c}}
		{\frac{{\delta {t^4}}}{4}}&{\frac{{\delta {t^3}}}{2}}\\
		{\frac{{\delta {t^3}}}{2}}&{\delta {t^2}}
\end{array}} \right],
\end{align}
where ${I_3}$ represents the $3\times 3$ unit matrix, $\sigma _v^2 = 5m{s^{ - 2}}$, $\otimes$ represents the Kronecker product and $\delta t = 1s$ denotes the sampling
interval. The plane consists of three motion modes i.e. $\mathbb{M}(c_p)=3$ with the linear motion, a counterclockwise turn rate of $-4^{\circ}/s$ and
a clockwise turn rate of $4^{\circ} /s$. For the plane, the notation $\zeta$ equals to $-4\pi/180$ or $4\pi/180$. The UAV consists of three motion modes i.e. $\mathbb{M}(c_u)=3$ with the linear motion, a counterclockwise turn rate of $-15^{\circ}/s$ and
a clockwise turn rate of $15^{\circ} /s$. For the UAV, the notation $\zeta$ equals to $-15\pi/180$ or $15\pi/180$. The surviving probabilities of the plane and UAV are given as ${p_{S}(c_p)} ={p_{S}(c_u)} =0.99$. The detection probabilities of the plane and UAV are given as ${p_{D}(c_p)} =0.99,~{p_{D}(c_u)} =0.95$, respectively. The number of clutter per scan is Poisson distributed with mean $\lambda_c =30$. We assume that surviving probabilities, detection probabilities and clutter rate are given as a prior knowledge. The initial models of targets are shown in Table \ref{Target States} and the death time here refers to the last time a target existing. 
\begin{figure}[!t]
	\centering
	\includegraphics[width=3in]{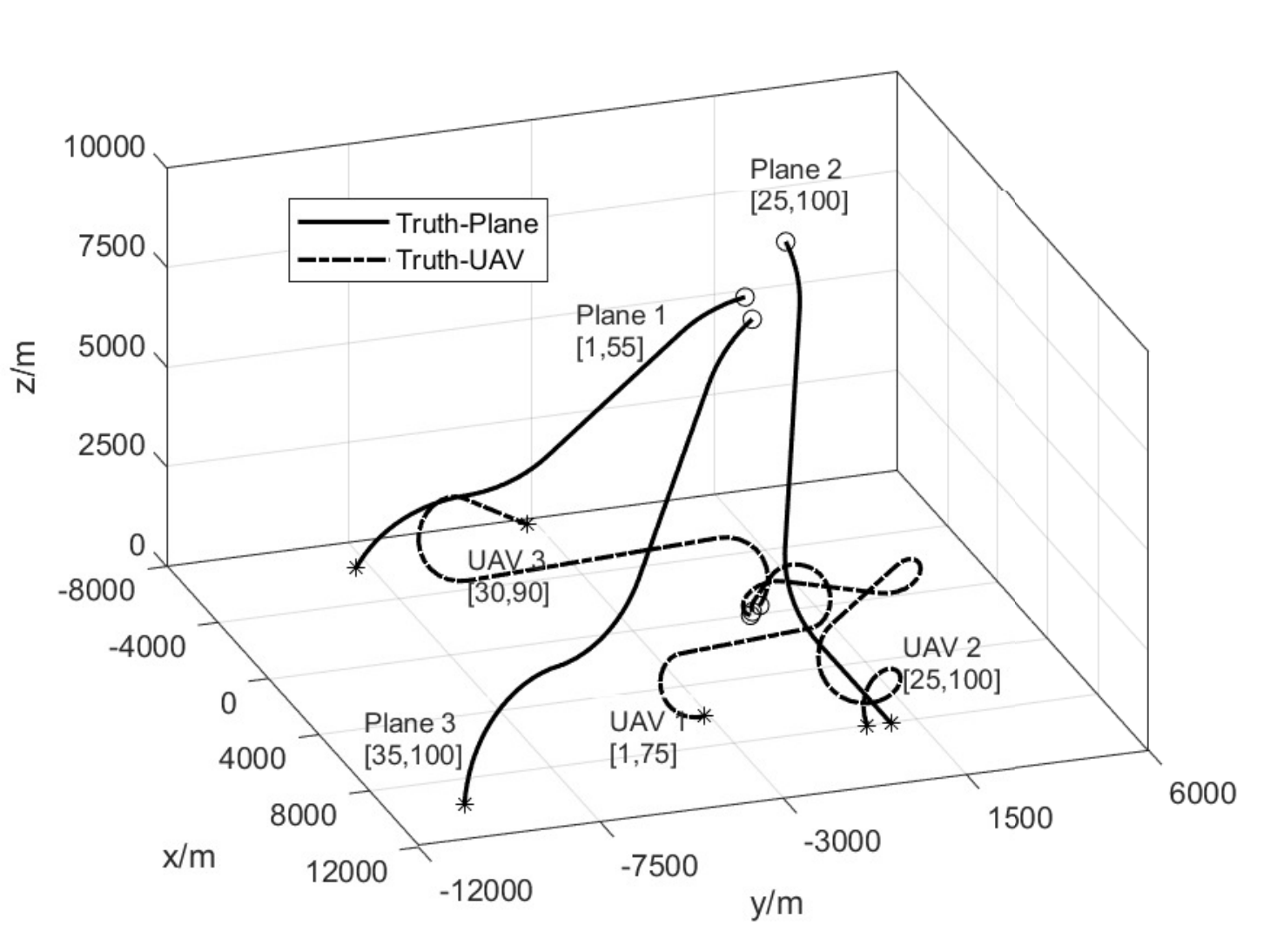}
	\caption{The true trajectory of the Planes and UAVs in 100s. The start and end
		points for each trajectory are marked by $\text{o}$ and $*$, respectively. The targets classes, as well as their birth and death time are marked in this Fig.}
	\label{JTC-state}
\end{figure}
\begin{figure}[!t]
	\centering
	\includegraphics[width=3in]{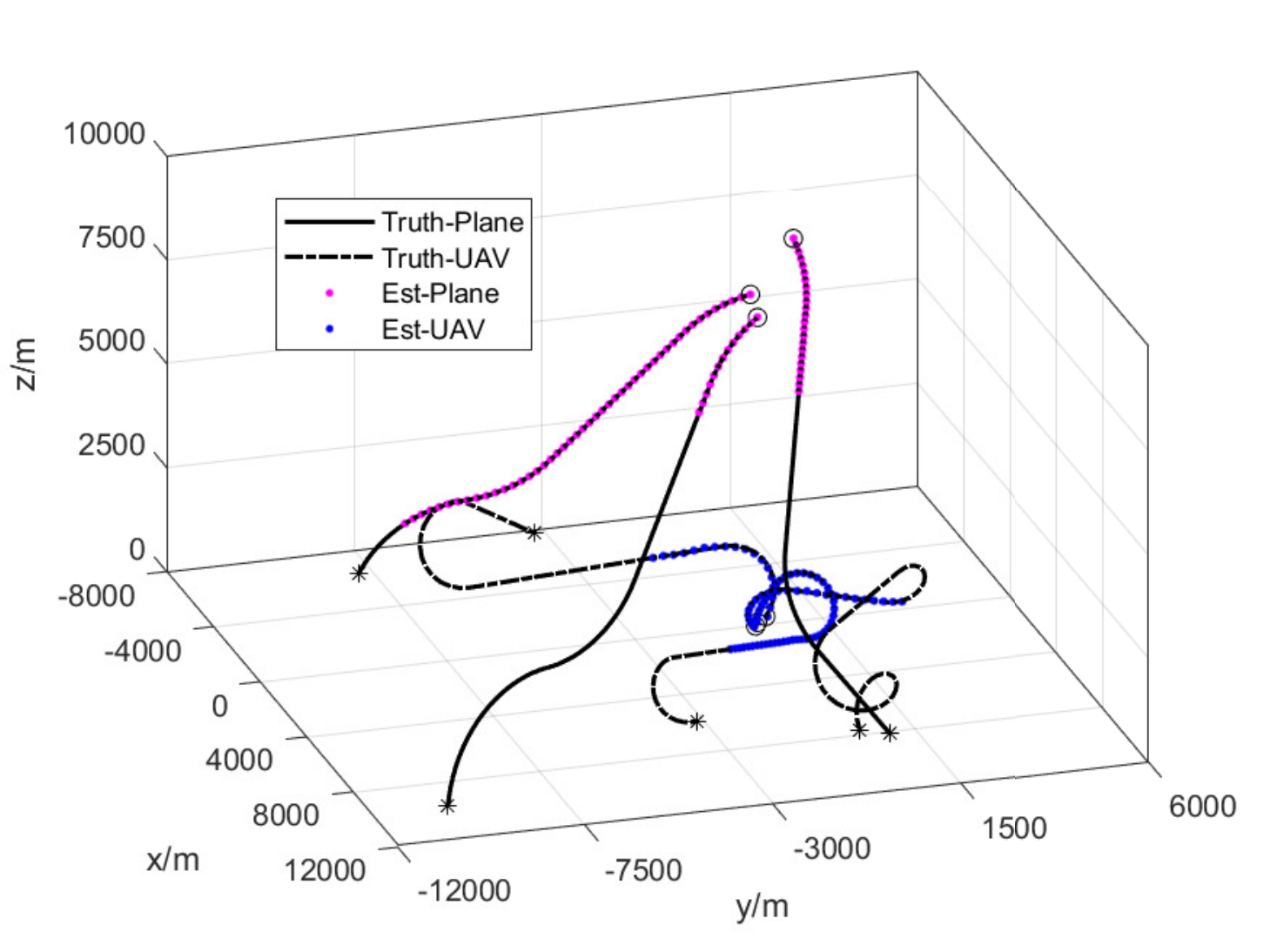}
	\caption{The trajectory estimation and classification results of the GM-JTC-TPHD filter at time $k=47$. }
	\label{JTC-state-47}
\end{figure}
\par Besides, the birth process is Poisson with parameters ${J^k_\gamma }=4$, ${\omega^k_{\gamma_c} }=0.03\times[0.5,0.5]$, ${\omega^k_{\gamma_r} }=[0.3,0.35,0.35]$ and ${\widehat P^k_\gamma }=\text{diag}([50,50,50,50,50,50])^2$. For each $j \in \left\{ {1,2,3,4} \right\},~\widehat m_{\gamma,1}^k=[0, 0, 0, 0, 8000, 0]^\top,~\widehat m_{\gamma,2}^k=[-1000, 0, 1000, 0, 9000, 0]^\top$, $~\widehat m_{\gamma,3}^k=[0, 0, 100, 0, 7500$, $0]^\top$,$~\widehat m_{\gamma,4}^k=[50, 0, 50, 0, 0, 0]^\top$
The value of the $L$-scan approximation is set as $L=5$. The switching between different motion modes in both classes is taken as the same, which is given by the following
Markovian model transition probability matrices
\begin{align}
		f_r(r^i|r^{i-1}) =& \left[ {\begin{array}{*{20}{c}}
			{0.8}&{0.1}&{0.1}\\
			{0.1}&{0.8}&{0.1}\\
			{0.1}&{0.1}&{0.8}
	\end{array}} \right],
\end{align}
\par In pruning and absorption, the threshold of weight is 
$\Gamma_p=10^{-5}$, the threshold of absorption is given as $\Gamma_a=4$ and the maximum is limited as $J_{max}=50$. The trajectory metric error (TM) \cite{Angel2020TM} with parameters $p = 2, c = 100, \gamma  = 1$ is used to characterize the error between the estimated and
truth for the GM-JTC-TPHD filter. By running 1000 Monte Carlo experiments, the performance of the GM-JTC-TPHD filter is given as follows. 
\begin{table}[!t]
	\centering
	\caption{The Initial Target States}
		\scriptsize
	\label{Target States}
	\begin{tabular}{c|c|c|c}
		\hline
		&Birth State&Birth Time&Death Time\\
		\hline
		Plane 1&$\left[0, -100, 0, -200, 8000, -60\right]^{\top}$&1&55\\	 \hline
		Plane 2&$\left[-1000, 120, 1000, 50, 9000, -120 \right]^{\top}$&25&100\\	 \hline
		Plane 3&$\left[0, 150, 100, -200, 7500, -40\right]^{\top}$&35&100\\
		\hline 
		UAV 1&$\left[0, 110, 0, -100, 0, 100\right]^{\top}$&1&75\\
		\hline
		UAV 2&$\left[100, 150, 0, -180, 10, 50\right]^{\top}$&25&100\\
		\hline
		UAV 3&$\left[ -100, -90, 10, 200, 0, 100\right]^{\top}$&30&90\\
		\hline
	\end{tabular}
\end{table}
\begin{figure}[!t]
	\centering
	\includegraphics[width=2.8in]{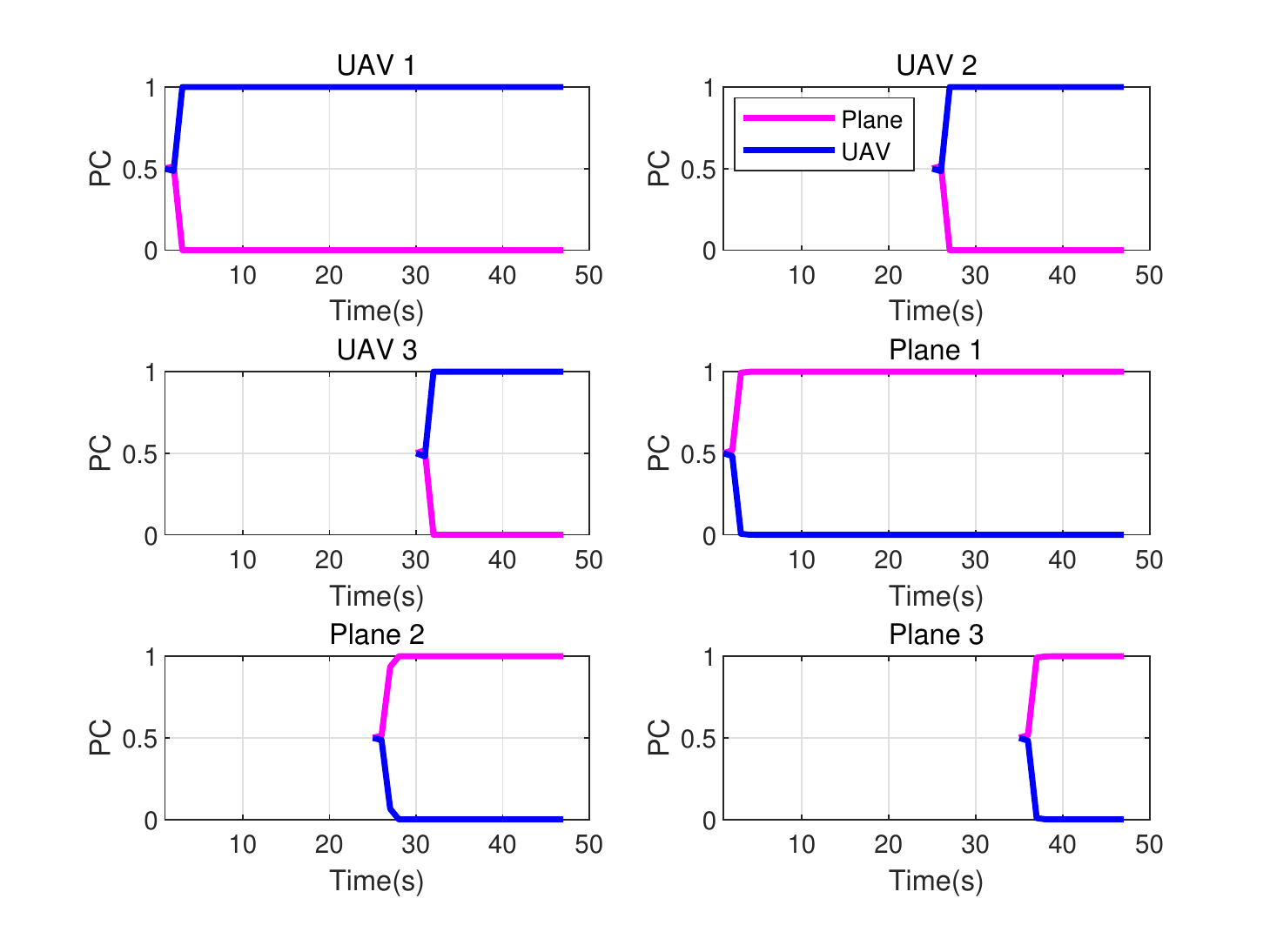}
	\caption{It shows the classification histories of trajectories in Fig. \ref{JTC-state-47} with one Monte Carlo experiment, where the word PC is the abbreviation of the probability of classes. The classification results of targets of a single frame time is the class with the maximum probability at this moment. }
	\label{Classificaiton}
\end{figure}

\begin{figure}[!t]
	\centering
	\includegraphics[width=3in]{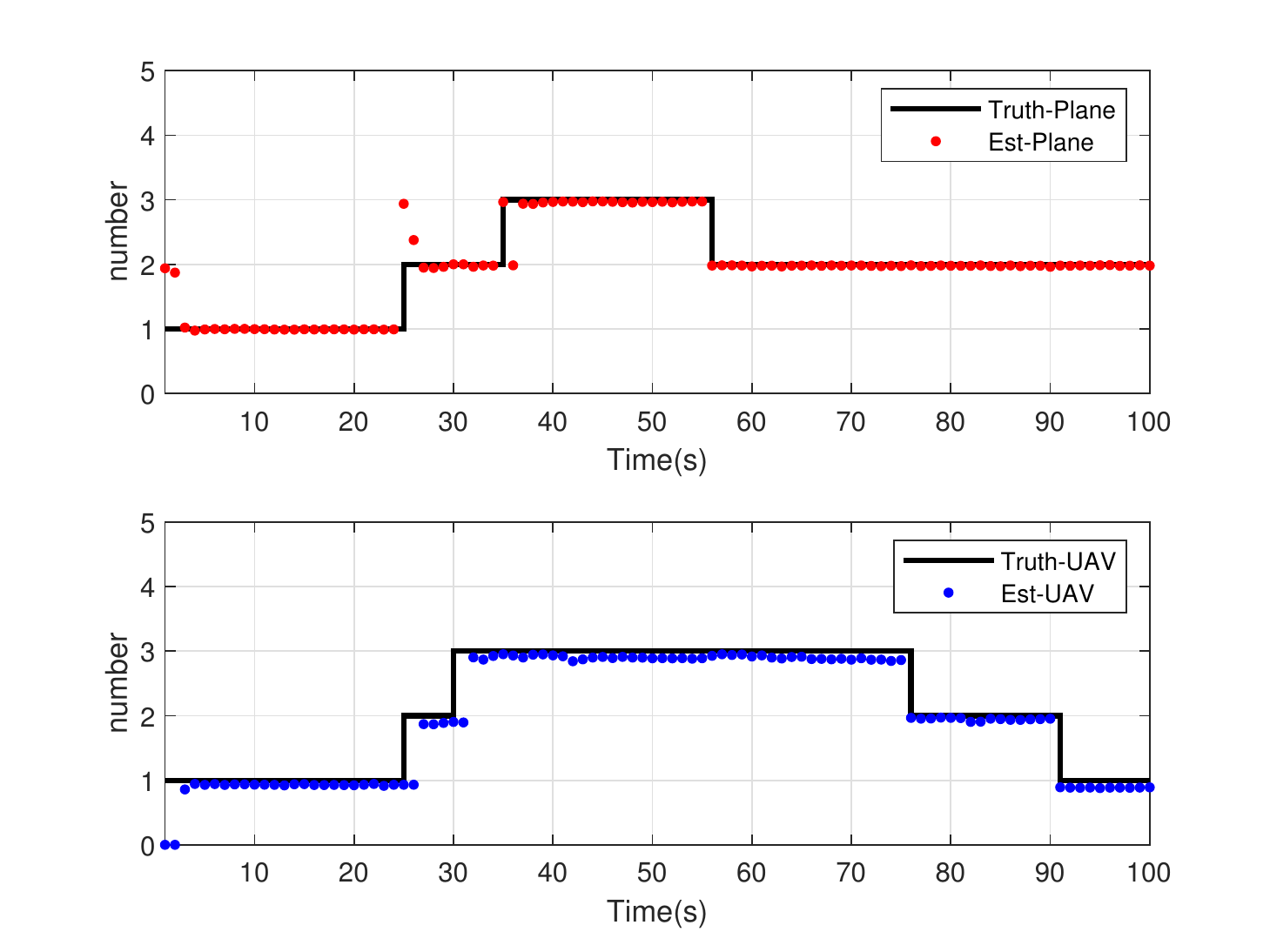}
	\caption{The average estimation of the target number of two classes in the GM-JTC-TPHD filter with multiple Monte Carlo experiments.}
	\label{JTC-number}
\end{figure}
\begin{figure}[!t]
	\centering
	\includegraphics[width=2.8in]{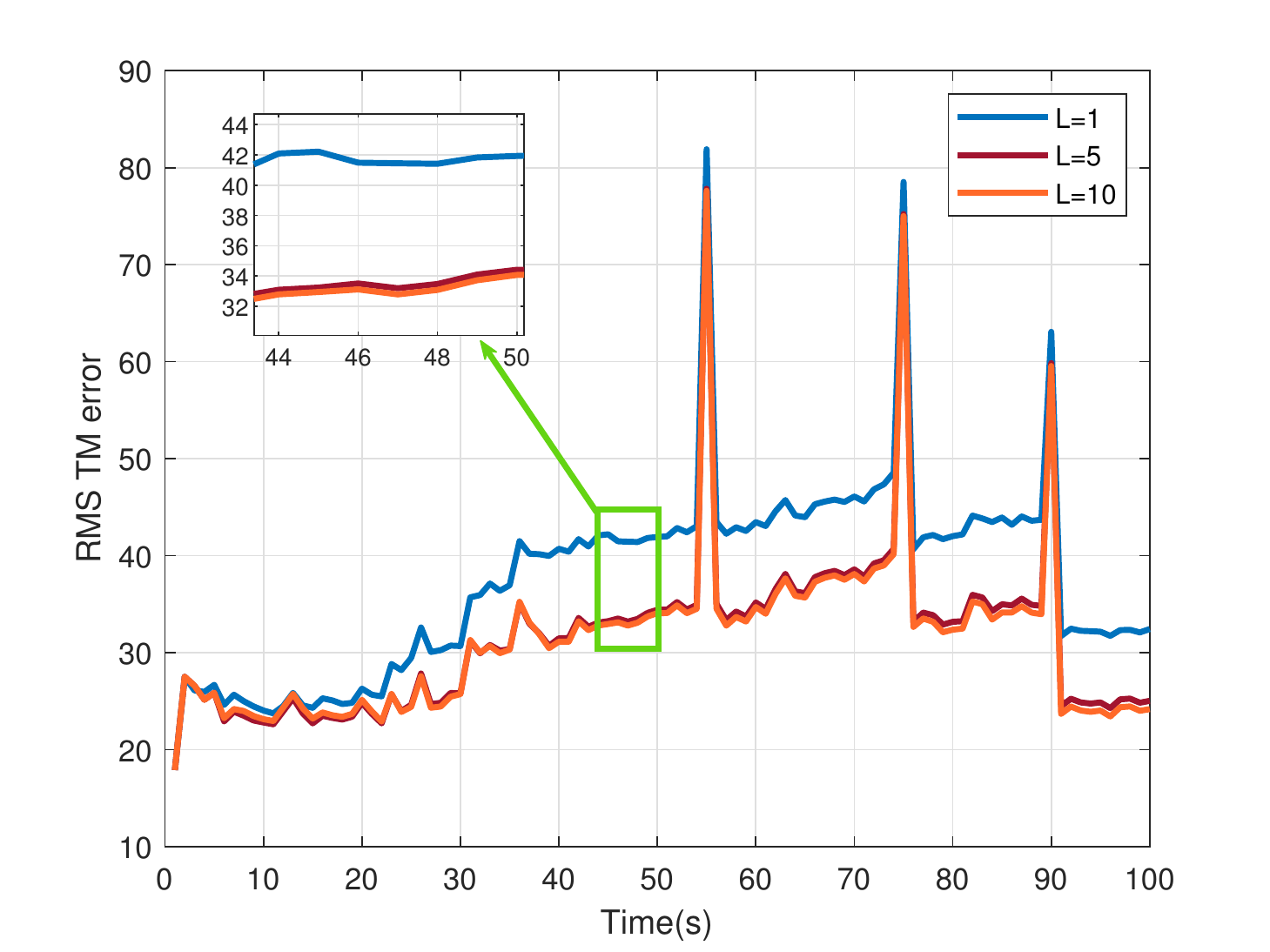}
	\caption{ The RMS trajectory metric error of all alive trajectories with different $L$, considering both classes. }
	\label{Lscan}
\end{figure}
\par It can be seen from Fig. \ref{JTC-state-47}
that the GM-JTC-TPHD filter can achieve excellent performance in the estimation of the trajectory states and classification results. 
Then, it can be seen from Figs.\ref{Classificaiton} and \ref{JTC-number} that the GM-JTC-TPHD filter can correctly classify the plane and UAV with time progressing, but performs fluctuation with the changes of the number
of targets. Finally, the influence of different values of the $L$-scan approximation is shown in Fig. \ref{Lscan}. As expect, an increasing $L$
can improve estimation performance and reduce the errors, while the improvement becomes less and less with increasing $L$. In addition, The averaged times to run one Monte Carlo iteration with a 2.8 GHz Intel i7 laptop are given as
: 8.69 seconds, 9.21 seconds, 15.57 seconds and 28.13 seconds, with $L \in \{1, 5, 10, 30\}$. It is worth nothing that, when $L=1$, the GM-JTC-TPHD filter is equivalent to the GM-JTD-PHD filter \cite{Yw2012JTC} in Fig. \ref{Lscan}. In addition, considering both the computational efficiency and performance, $L=5$ is a suitable value in this scenario.

\section{Conclusion}
In this paper, the recursion of the JTC-TPHD filter is derived, which can not only estimate trajectories, but also classify different kinds of targets. The Gaussian mixture implementation is presented for the JTC-TPHD filter, which is referred to as the GM-JTC-TPHD filter. The $L$-scan approximation is also applied to the GM-JTC-TPHD filter to achieve a fast implementation. Simulation results demonstrate that the GM-JTC-TPHD filter can achieve excellent performance in both tracking and classification. However, this paper only considers the classification based on the target state and motion models of a single frame time. Future works will research the special association between trajectories and classes to obtain a better classification results.
\bibliographystyle{IEEEtran}
\bibliography{UN}

\begin{thebibliography}{10}
\providecommand{\url}[1]{#1}
\csname url@samestyle\endcsname
\providecommand{\newblock}{\relax}
\providecommand{\bibinfo}[2]{#2}
\providecommand{\BIBentrySTDinterwordspacing}{\spaceskip=0pt\relax}
\providecommand{\BIBentryALTinterwordstretchfactor}{4}
\providecommand{\BIBentryALTinterwordspacing}{\spaceskip=\fontdimen2\font plus
\BIBentryALTinterwordstretchfactor\fontdimen3\font minus
  \fontdimen4\font\relax}
\providecommand{\BIBforeignlanguage}[2]{{%
\expandafter\ifx\csname l@#1\endcsname\relax
\typeout{** WARNING: IEEEtran.bst: No hyphenation pattern has been}%
\typeout{** loaded for the language `#1'. Using the pattern for}%
\typeout{** the default language instead.}%
\else
\language=\csname l@#1\endcsname
\fi
#2}}
\providecommand{\BIBdecl}{\relax}
\BIBdecl

\bibitem{Mahler2007RFSbook}
R.~Mahler, \emph{Statistical Multisource Multitarget Information Fusion}.\hskip
  1em plus 0.5em minus 0.4em\relax Norwood, MA, USA: Artech House, 2007.

\bibitem{Mahler2003RFS}
------, ``Multi-target {B}ayes filter via first-order multi-target moments,''
  \emph{IEEE Trans. Aerosp. Electron. Syst.}, vol.~39, no.~4, pp. 1152--1178,
  Oct. 2003.

\bibitem{Vo2006PHD}
B.-N. Vo and W.-K. Ma, ``The {G}aussian mixture probability hypothesis density
  filter,'' \emph{IEEE Trans. Signal Process.}, vol.~54, no.~11, pp.
  4091--4104, Nov. 2006.

\bibitem{Angel2015KLD}
A.~F. García-Fernández and B.-N. Vo, ``Derivation of the {PHD} and {CPHD}
  filters based on direct {K}ullback-{L}eibler divergence minimization,''
  \emph{IEEE Trans. Signal Process.}, vol.~63, no.~21, pp. 5812--5820, Nov.
  2015.

\bibitem{Vo2005smc}
B.-N. Vo, S.~Singh, and A.~Doucet, ``Sequential {M}onte {C}arlo methods for
  multi-target filtering with random finite sets,'' \emph{IEEE Trans.
  Aerosp.Electron. Syst.}, vol.~41, no.~4, pp. 1224--1245, 2005.

\bibitem{Vo2007CPHD}
B.-N. Vo and A.~Cantoni, ``Analytic implementations of the cardinalized
  probability hypothesis density filter,'' \emph{IEEE Trans. Signal Process.},
  vol.~55, no.~7, pp. 3553--3567, Jul. 2007.

\bibitem{Vo2009MB}
B.-T. Vo, B.~N. Vo, and A.~Cantoni, ``The cardinality balanced multi-target
  multi-{B}ernoulli filter and its implementations,'' \emph{IEEE Trans. Signal
  Process.}, vol.~57, no.~2, pp. 409--423, 2009.

\bibitem{Angel2018PMBM}
A.~F. García-Fernández, J.~L. Williams, K.~Granström, and L.~Svensson,
  ``Poisson multi-{B}ernoulli mixture filter: direct derivation and
  implementation,'' \emph{IEEE Trans. Aerosp.Electron. Syst.}, vol.~54, no.~4,
  pp. 1883--1901, Aug. 2018.

\bibitem{Vo2013GLMB}
B.-T. Vo and B.-N. Vo, ``Labeled {R}andom {F}inite {S}ets and {M}ulti-{O}bject
  {C}onjugate {P}riors,'' \emph{IEEE Trans. Signal Process.}, vol.~61, no.~13,
  July. 2013.

\bibitem{Vo2014GLMB}
B.-T. Vo, B.-N. Vo, and D.~Phung, ``{L}abeled {R}andom {F}inite {S}ets and the
  {B}ayes {M}ulti-{T}arget {T}racking {F}ilter,'' \emph{IEEE Trans. Signal
  Process.}, vol.~62, no.~24, pp. 6554--6567, Dec. 2014.

\bibitem{Reuter2014LMB}
S.~Reuter, B.-T. Vo, B.-N. Vo, and K.~DieTMayer, ``The {L}abeled
  {M}ulti-{B}ernoulli {F}ilter,'' \emph{IEEE Trans. Signal Process.}, vol.~62,
  no.~12, pp. 3246--3260, June. 2014.

\bibitem{Vo2019MsGLMB}
B.-N. Vo and B.-T. Vo, ``A multi-scan labeled random finite set model for
  multi-object state estimation,'' \emph{IEEE Trans. Signal Process.}, vol.~67,
  no.~19, pp. 1003--1016, Oct. 2019.

\bibitem{Angel2019TPHD}
A.~F. García-Fernández and L.~Svensson, ``Trajectory {PHD} and {CPHD}
  filters,'' \emph{IEEE Trans. Signal Process.}, vol.~67, no.~22, pp.
  1003--1016, Nov. 2019.

\bibitem{Granstrom2018TPMBM}
K.~Granström, L.~Svensson, Y.~Xia, and J.~L. Williams, ``Poisson
  multi-{B}ernoulli mixture trackers: {C}ontinuity through random finite sets
  of trajectories,'' \emph{21st International Conference on Information Fusion
  (FUSION)}, pp. 973--981, 2018.

\bibitem{Angel2020TMB}
A.~F. García-Fernández, L.~Svensson, J.~L. Williams, Y.~Xia, and
  K.~Granström, ``Trajectory multi-{B}ernoulli filters for multi-target
  tracking based on sets of trajectories,'' \emph{23rd International Conference
  on Information Fusion (FUSION)}, pp. 1003--1016, 2020.

\bibitem{Angel2020TMTT}
A.~F. García-Fernández, L.~Svensson, and M.~R. Morelande, ``Multiple target
  tracking based on sets of trajectories,'' \emph{IEEE Trans. Aerosp. Electron.
  Syst.}, vol.~56, no.~3, pp. 1685--1707, Jun. 2020.

\bibitem{Yw2012JTC}
Y.~Wei, F.~Yaowen, L.~Jianqian, and L.~Xiang, ``Joint {D}etection, {T}racking,
  and {C}lassification of {M}ultiple {T}argets in {C}lutter {U}sing the {PHD}
  filter,'' \emph{IEEE Trans. Aerosp. Electron. Syst.}, vol.~48, p.
  3594–3608, 2012.

\bibitem{Magnant2016MSPHD}
C.~Magnant, A.~Giremus, E.~Grivel, L.~Ratton, and B.~Joseph, ``Multi-target
  tracking using a {PHD}-based joint tracking and classification algorithm,''
  \emph{2016 IEEE Radar Conference (RadarConf)}, pp. 1--6, 2016.

\bibitem{Lsun2014JTC}
L.~Sun, J.~Lan, and X.~R. Li, ``Joint tracking and classification of extended
  object based on support functions,'' \emph{17th International Conference on
  Information Fusion (FUSION)}, pp. 1--8, 2014.

\bibitem{Challa2001JTC-ESM}
S.~Challa and G.~W. Pulford, ``Joint target tracking and classification using
  radar and esm sensors,'' \emph{IEEE Trans. Aerosp. Electron. Syst.}, vol.~37,
  no.~3, p. 1039–1055, 2001.

\bibitem{JTC-GLMB1}
D.~Chen, C.~Li, and H.~Ji, ``Multi-target joint detection, tracking and
  classification with merged measurements using generalized labeled
  multi-{B}ernoulli filter,'' \emph{20th International Conference on
  Information Fusion (FUSION)}, pp. 1--8, 2017.

\bibitem{JTC-GLMB2}
M.~Li, Z.~Jing, P.~Dong, and H.~Pan, ``Multi-target joint detection, tracking
  and classification using generalized labeled {M}ulti-{B}ernoulli filter with
  {B}ayes risk,'' \emph{in Proc. 19th Int. Conf. Inf. Fusion}, pp. 680--687,
  2016.

\bibitem{JTC1}
W.~Cao, J.~Lan, and X.~R. Li, ``Joint tracking and classification based on
  recursive joint decision and estimation using multi-sensor data,'' \emph{14th
  International Conference on Information Fusion (FUSION)}, pp. 1--8, 2011.

\bibitem{Pasha2009JMS}
S.~A. Pasha, H.~D.~T. B.~Vo, and W.~Ma, ``A {G}aussian {M}ixture {PHD} {F}ilter
  for {J}ump {M}arkov {S}ystem {M}odels,'' \emph{IEEE Trans. Aerosp. Electron.
  Syst.}, vol.~45, no.~3, pp. 919--936, 2009.

\bibitem{MM1}
S.~Reuter, A.~Scheel, and K.~Dietmayer, ``The {M}ultiple {M}odel {L}abeled
  {M}ulti-{B}ernoulli {F}ilter,'' \emph{18th International Conference on
  Information Fusion (FUSION)}, pp. 1574--1580, 2015.

\bibitem{MM2}
R.~Georgescu and P.~Willett, ``The {M}ultiple {M}odel {CPHD} {T}racker,''
  \emph{IEEE Trans. Signal Process.}, vol.~60, no.~4, pp. 1741--1751, 2012.

\bibitem{MM3}
W.~Yi, M.~Jiang, and R.~Hoseinnezhad, ``The {M}ultiple {M}odel {V}o–{V}o
  {F}ilter,'' \emph{IEEE Trans. Aerosp. Electron. Syst.}, vol.~53, no.~2, pp.
  1045--1054, April 2017.

\bibitem{CJDE1}
X.~R. Li, ``Optimal bayes joint decision and estimation,'' \emph{10th
  International Conference on Information Fusion (FUSION)}, pp. 1--8, 2007.

\bibitem{CJDE2}
Y.~Liu and X.~R. Li, ``Recursive {J}oint {D}ecision and {E}stimation {B}ased on
  {G}eneralized {B}ayes {R}isk,'' \emph{14th International Conference on
  Information Fusion (FUSION)}, pp. 2066--2073, 2011.

\bibitem{JMS-TPHD}
\BIBentryALTinterwordspacing
B.~Zhang and W.~Yi, ``The {T}rajectory {PHD} {F}ilter for {J}ump {M}arkov
  {S}ystem {M}odels and {I}ts {G}aussian {M}ixture {I}mplementation,'' 2021.
  [Online]. Available: \url{https://arxiv.org/pdf/2008.03914.pdf.}
\BIBentrySTDinterwordspacing

\bibitem{KF-1970}
A.~H. Jazwinski, \emph{tochastic Processes and Filtering Theory}.\hskip 1em
  plus 0.5em minus 0.4em\relax New York: Academic, 1970.

\bibitem{Angel2020TM}
A.~F. García-Fernández, A.~S. Rahmathullah, and L.~Svensson, ``A metric on
  the space of finite sets of trajectories for evaluation of multi-target
  tracking algorithms,'' \emph{IEEE Trans. Signal Process.}, vol.~68, pp.
  3917--3928, 2020.

\end{thebibliography}

\end{document}